\documentclass[aps,pre,twocolumn,superscriptaddress,showpacs,showkeys ]{revtex4-1}

\usepackage{graphicx}                   
\usepackage{hyperref}                   
\usepackage{amsmath,amssymb}            
\usepackage{epstopdf}
\usepackage{subcaption}					

\usepackage{color}
\definecolor{orange}{rgb}{1,0.5,0}
\definecolor{brown}{rgb}{0.65, 0.16, 0.16}
\definecolor{phlox}{rgb}{0.87, 0.0, 1.0}

\begin{document}

\title{Correlation effects in the diluteness pattern in non-integral dimensional systems\\
	on $\nu=\frac{4}{5}$ superdiffusion process}

\author{J. Cheraghalizadeh}
\affiliation{Department of Physics, University of Mohaghegh Ardabili, P.O. Box 179, Ardabil, Iran}
\email{jafarcheraghalizadeh@gmail.com}

\author{M. N. Najafi*}
\affiliation{Department of Physics, University of Mohaghegh Ardabili, P.O. Box 179, Ardabil, Iran}
\email{morteza.nattagh@gmail.com}

\begin{abstract}
The effect of the correlations in the diluteness pattern in the systems with non-integral dimensionality, on $\nu=\frac{4}{5}$ superdiffusion process is considered in this paper. These spatial correlations have proved to be very effective in the critical phenomena. To simulate the particles motion in this process, we employ the loop erased random walk (LERW). The spatial correlations between imperfections (site-diluteness) also have been modeled by the Ising model on a square lattice. It models the forbidden regions into which the particles are not allowed to enter. The correlations are controlled by an artificial temperature $T$. The trace of the walkers is shown to be self-similar, whose fingerprint is the power-law behaviors. The detailed analysis of the random walker's traces reveal that the (Ising-type) correlations affect their geometrical properties. At the critical artificial temperature $T_c$ we observe that the exponent of end-to-end distance $\nu$ becomes $0.807\pm 0.002$. The fractal dimension of the walker's trace is the other geometrical quantity which scales inversely with the square root of the correlation length of the Ising model, i.e. $D_f(T)-D_f(T_c)\sim \zeta^{-\alpha}$ in which $\alpha=0.43\pm 0.05$. The winding angle test also reveals that the traces are compatible with the Schramm-Loewner evolution theory, and the diffusivity parameter for $T=T_c$ is $\kappa=1.89\pm 0.05$.
\end{abstract}


\maketitle

\section{Introduction}
The problem of superdiffusion and subdiffusion processes, and the corresponding correlated random walks, has received attention within the literature to describe many physical scenarios, mostly for crowded systems~\cite{metzler2000random,bouchaud1990anomalous}. Superdiffusion process in the protein diffusion within cells and also the diffusion in the porous media are important examples. This problem involves large number of phenomenon, ranging from heartbeat intervals to DNA sequences~\cite{buldyrev1994fractals}. Recently, anomalous diffusion was found in several systems including single particle movements in cytoplasm~\cite{regner2013anomalous}, worm-like micellar solutions~\cite{jeon2013anomalous}, telomeres in the nucleus of cells~\cite{bronstein2009transient}, ultra-cold atoms~\cite{sagi2012observation}. When the environmental disorder is present, this problem falls into the more general class, i.e. the critical phenomena on the imperfect and also fractal hosts~\cite{dhar1977lattices,dhar1978d,gefen1980critical}. This problem is realized simply by arresting the diffusing particles to enter some forbidden islands, which cause their diffusion properties change. For example by modeling the imperfections of the host media by the uncorrelated percolation theory, it is shown that the diffusion becomes anomalous ~\cite{havlin1987diffusion,bouchaud1990anomalous,sahimi2011flow}, and the fractal dimension ($D_f$) of the trace of particles considerably changes~\cite{kremer1981self,cheraghalizadeh2018self,daryaei2014loop,rammal1984self,chakrabarti1981statistics}. When the occupation probability of the host media $p$ is larger than the percolation threshold ($p_c$), two different scales are present: for small scales (compared with the correlation length of the percolation system) the exponents are anomalous, whereas for the large scales, the walkers behave in the same manner as the regular system~\cite{ben2000diffusion}. The problem is completely different for critical case ($p=p_c$) for which for all scales the anomalous exponents are obtained~\cite{daryaei2014loop}. \\
The anomalous diffusion problems can mostly be realized by some correlated random walks. Loop-erased random walk (LERW) is a correlated self-avoiding random trace which has deep relations with important statistical models. It is defined as the traces which are produced by random walkers who erase all of loops of the trace. It is well-known that in two-dimensional regular lattice, the fractal dimension of these traces is $\frac{5}{4}$, corresponding to $\nu=\frac{4}{5}$ diffusion process~\cite{Majumdar1992Exact} ($\nu$ being the end-to-end exponent, which is defined in the following sections). For simulating the $\nu=4/5$ diffusion process, one can consider LERW traces and test its properties, which is the purpose of the present paper. This model has relations to the uniform spanning trees~\cite{wilson1996symposium}, the BTW sandpiles~\cite{Majumdar1992Equivalence,najafi2012avalanche} and the $q$-state Potts model~\cite{Majumdar1992Exact}. The properties of the model have been well-studied on the regular lattices, e.g. its upper-critical dimension is $d_u=4$, and also in two dimensions, it is related to the Schramm-loewner evolution (SLE) with the diffusivity parameter $\kappa=2$, which corresponds to $c=-2$ conformal field theory. The ghost action of this universality class has elegantly been extracted by Bauer \textit{et al.}~\cite{bauer2008lerw}, who have also considered the problem in (mass-less ghosts) and out (massive ghosts) of criticality. Despite of this huge literature, a little attention has been paid to the response of the diffusing particles to the configuration of allowed areas of the host system. In~\cite{daryaei2014loop}, the authors revealed that loop erased random walkers create non-intersecting traces with $D_f\approx 1.22$ on the critical uncorrelated percolation lattice. In this problem the configuration of dilute sites is uncorrelated. One may investigate the same problem in the correlated percolation, to investigate the diffusion properties of the natural systems. In nature, the environmental disorder is commonly correlated, which is related to the process of the formation of host media~\cite{cheraghalizadeh2017mapping}. In this case, it is ideal to have a parameter that tunes the correlation and also the fraction of active sites. This is the aim of the present paper. We bring the correlations into the calculations by means of the Ising model with an artificial temperature $T$, which tunes the correlations between (quenched) environmental imperfections in the host media in which the loop-erased random walkers take their steps. We show that these correlations crucially change the properties of the model with respect to the uncorrelated percolation system.\\
The paper has been organized as follows: In the following section, we motivate this study and introduce and describe the model. The numerical methods and the results are presented in the section~\ref{NUMDet} which contains critical, as well as off-critical results. We end the paper by a conclusion.

\section{The construction of the problem}
\label{sec:model}
\begin{figure}
	\centerline{\includegraphics[scale=.35]{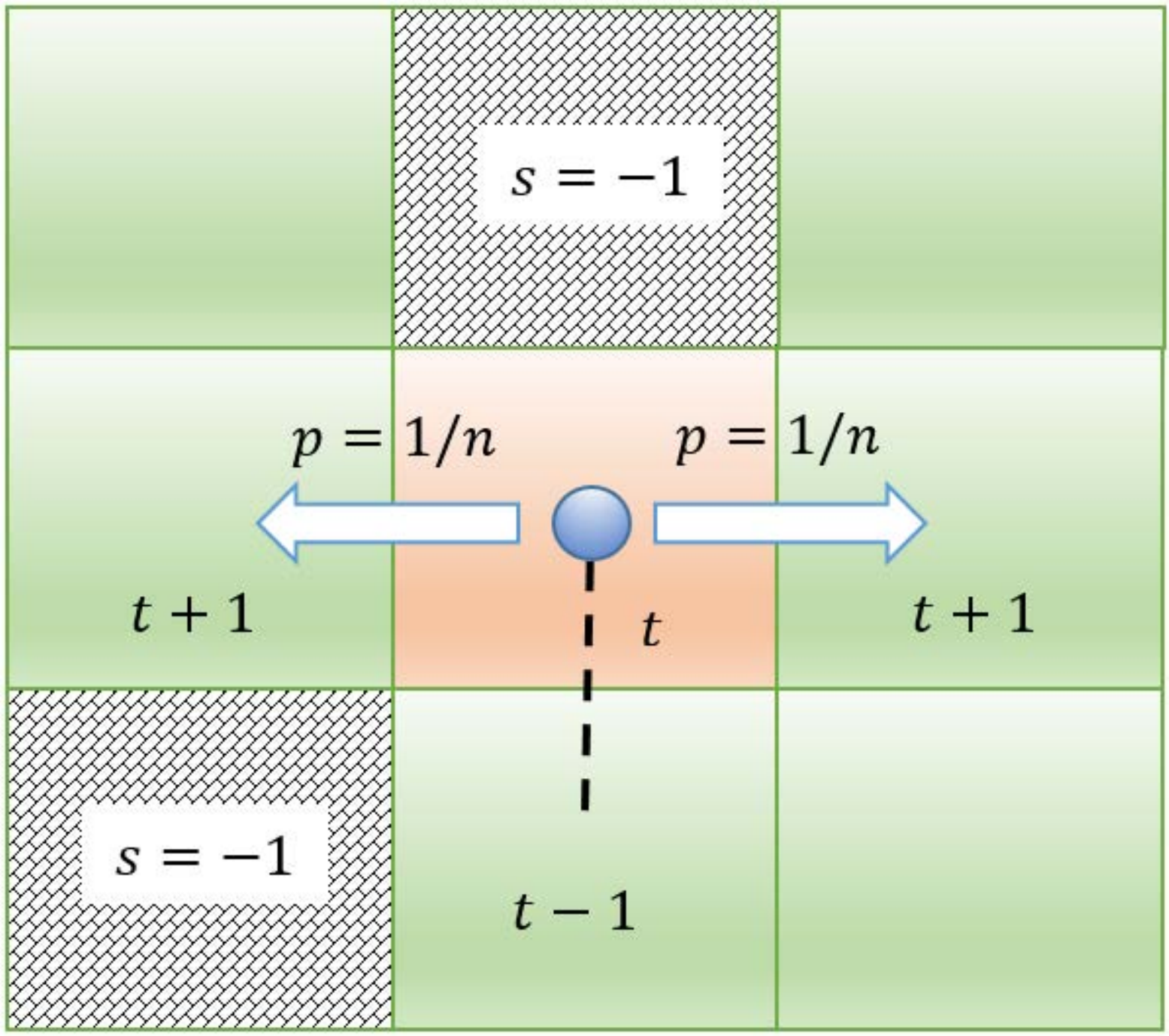}}
	\caption{A schematic set up of the LERW on site-diluted lattice. The random walker reaches the central site, having $n=2$ possibilities to move in the next step (the front site in not active ($s(i,j+1)=-1$), and also the random walker cannot return to $(i,j-1)$ since it forms a loop that should be removed). Therefore, the random walker chooses to move right or left with the equal probabilities $p=1/2$.}
	\label{fig:Movement}
\end{figure}
\begin{figure*}
	\begin{subfigure}{0.43\textwidth}\includegraphics[width=\textwidth]{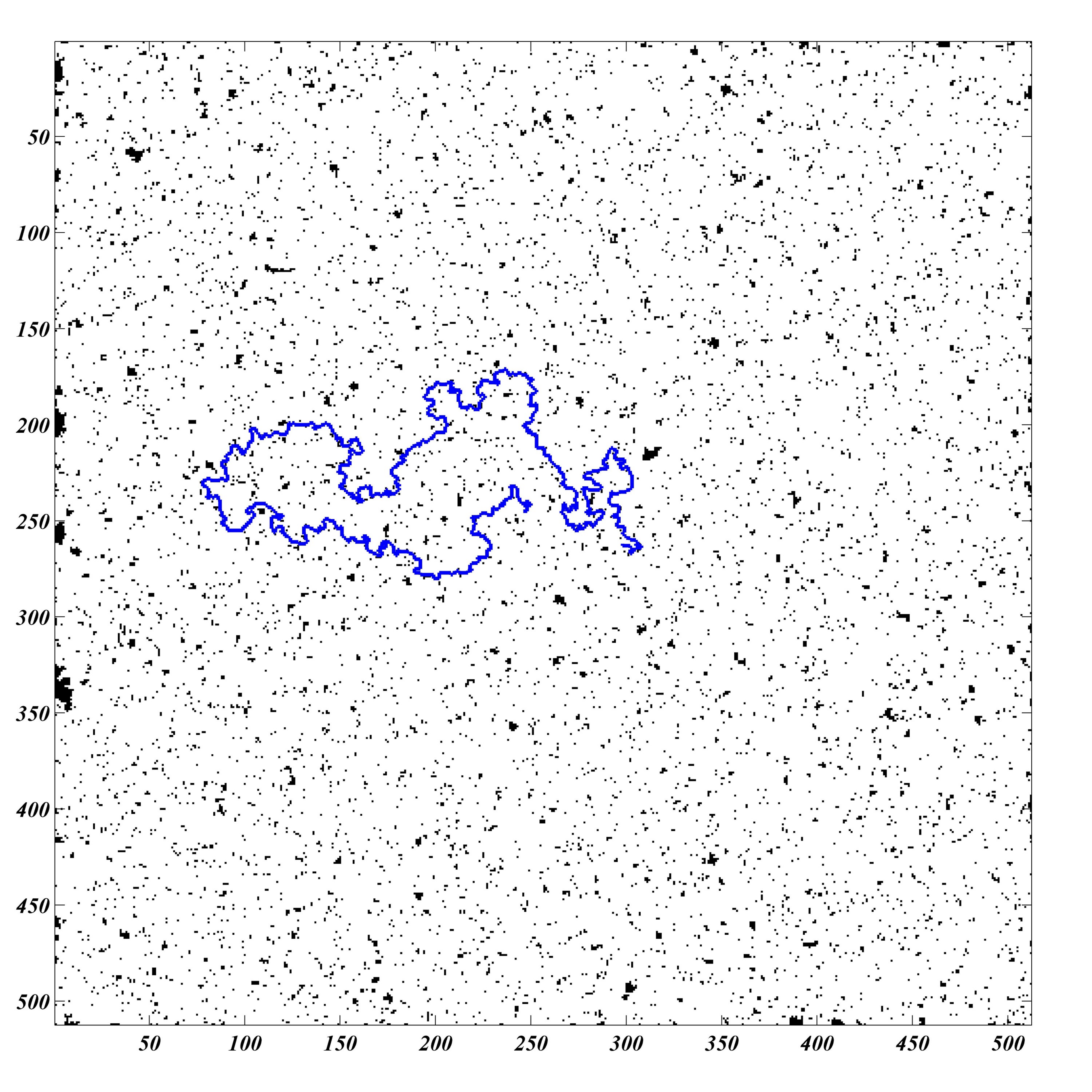}
		\caption{}
		\label{fig:sampleLowT}
	\end{subfigure}
	\centering
	\begin{subfigure}{0.43\textwidth}\includegraphics[width=\textwidth]{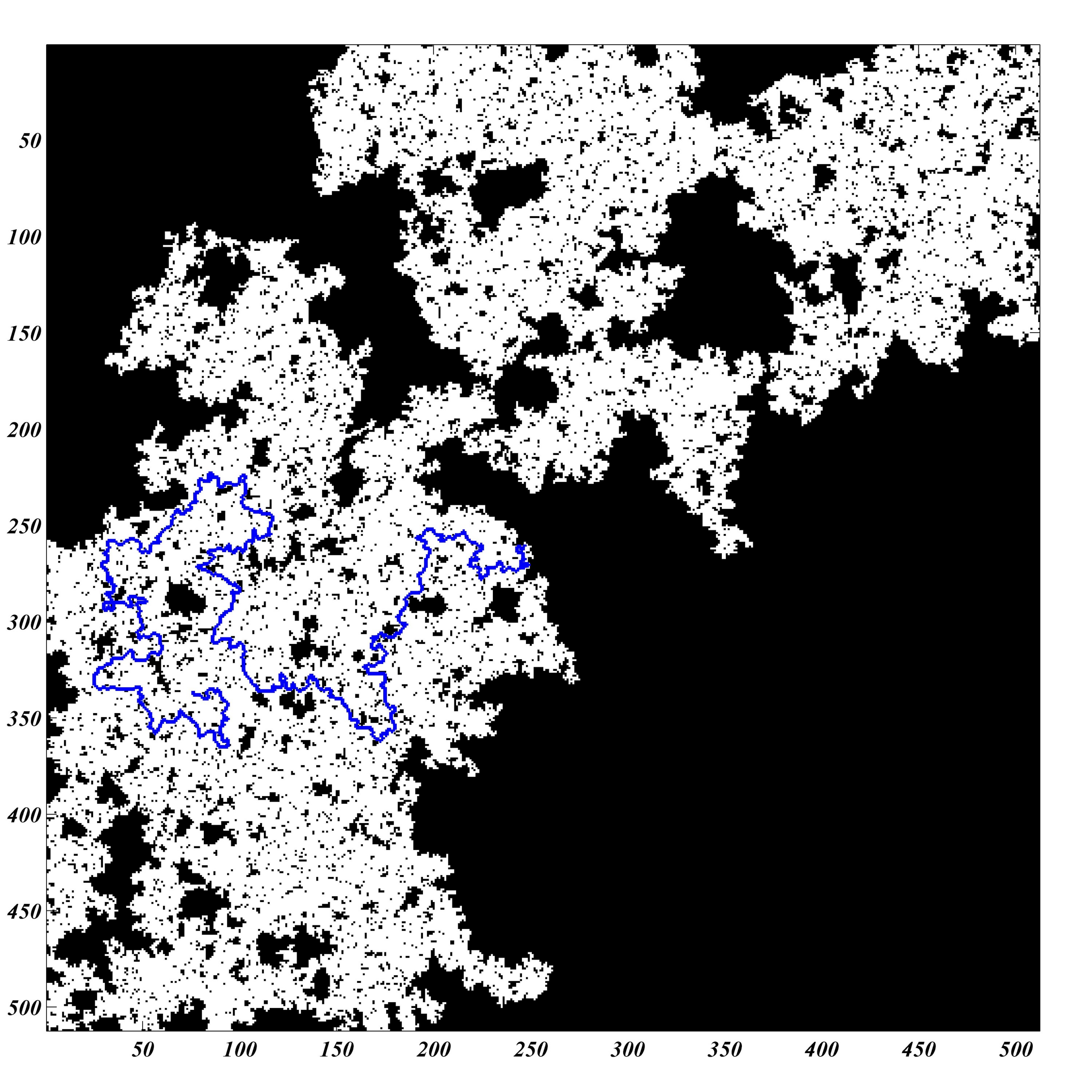}
		\caption{}
		\label{fig:sampleHighT}
	\end{subfigure}
	\begin{subfigure}{0.43\textwidth}\includegraphics[width=\textwidth]{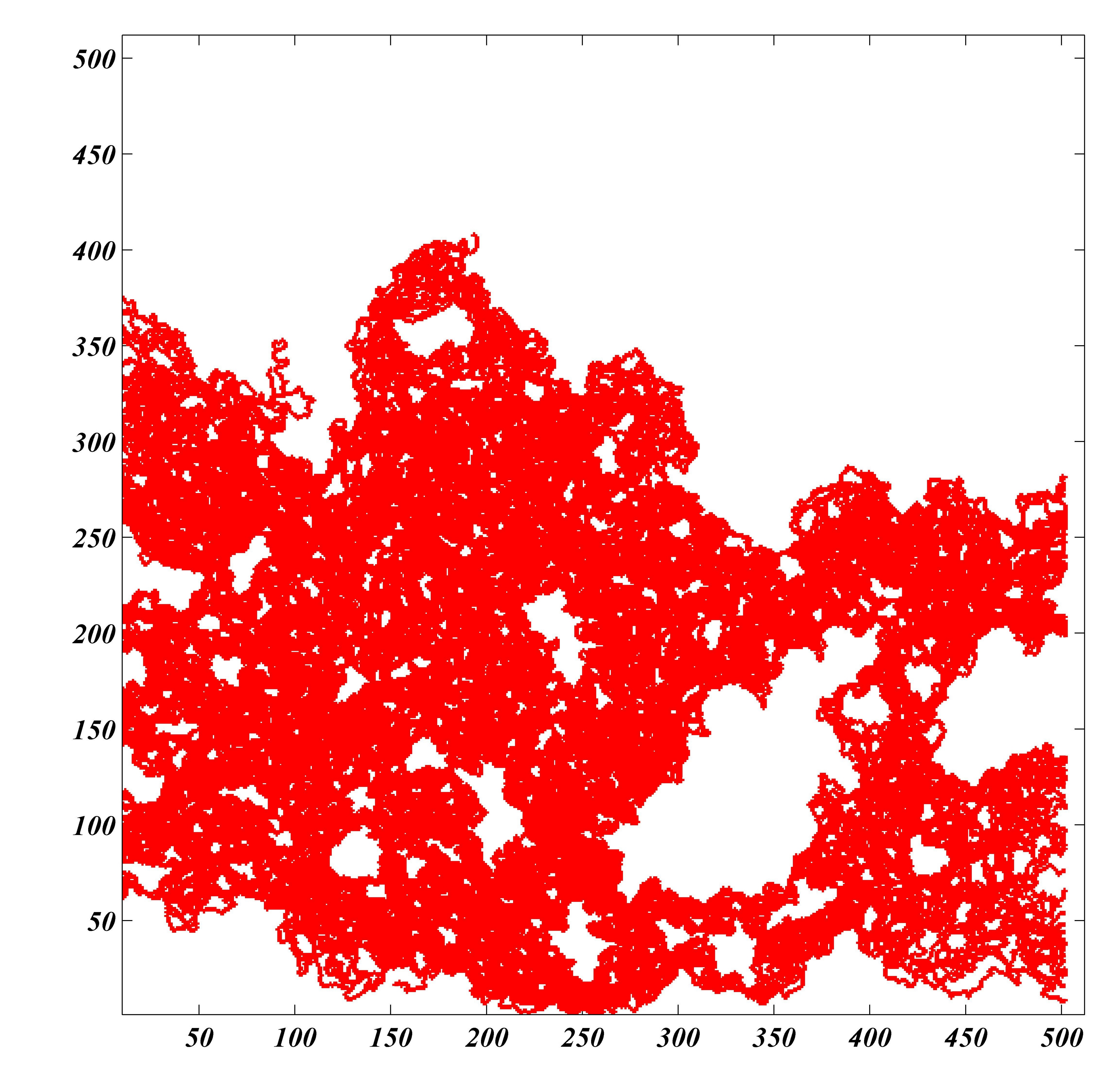}
		\caption{}
		\label{fig:500}
	\end{subfigure}
	\centering
	\begin{subfigure}{0.43\textwidth}\includegraphics[width=\textwidth]{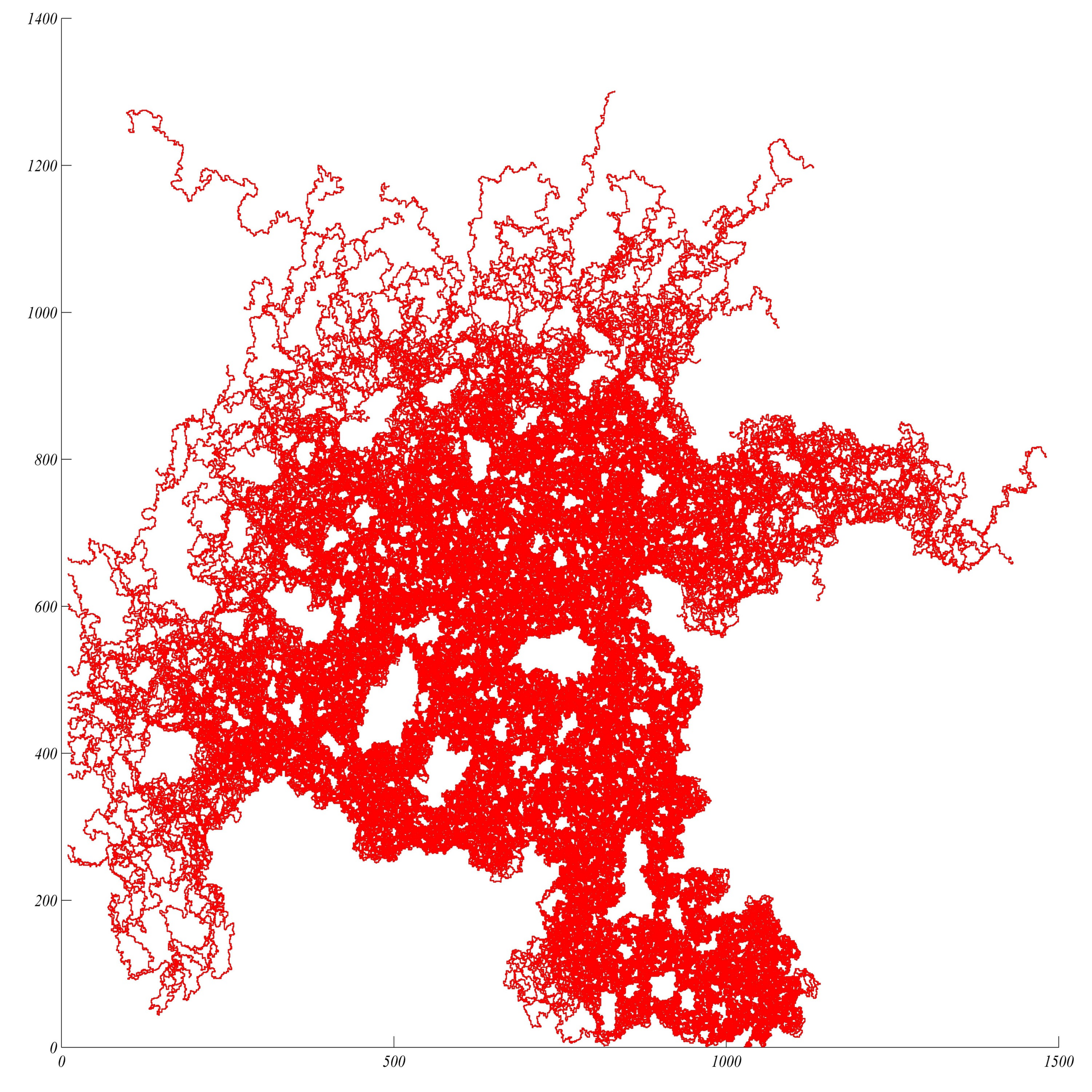}
		\caption{}
		\label{fig:500xy}
	\end{subfigure}
	\caption{(Color online): (a) An Ising sample with a loop erased random walker ($n=1500$ steps) for (a) $T=1.9\ll T_c$ (b) $T=T_c\approx 2.268$. The diffusion process: the particle traces in the background of Ising-correlated lattice for (c) $T=1.9$, and (d) $T=T_c$.}
	\label{IsingSamples}
\end{figure*}

Consider a diffusion process in which the mean squared displacement (MSD) $\left\langle R^2\right\rangle $ scales with time. In a typical diffusion process MSD scales linearly with time in any dimension. Anomalous diffusion is however a diffusion process with a non-linear relationship to time, i.e. $\left\langle R^2\right\rangle =D t^{2\nu}$, where $D$ is the diffusion coefficient, $\nu=0.5$ for ordinary diffusion process. If $\nu>0.5$ ($\nu<0.5$) the system is in the superdiffusion (subdiffusion) phase. \\
Diffusion processes can be realized by random walks, and the above mentioned non-linearity in time arises from the correlations in the random walks. Two important examples are self-avoiding walks (SAW), and loop erased random walks (LERW). The LERW is an important prototype of superdiffusion process which is analyzed in this paper. For this model $\nu=0.8$~\cite{lawler1999loop}.
LERW is defined as the trace of a random walker with removed loops. To be more precise, let $\left\lbrace \vec{r}_i\right\rbrace_{i=1}^{N}$ be a normal two-dimensional path of a random walker (which may include some loops). A loop with length $l$ is defined as the sub-path $\left\lbrace \vec{r}_{i_0+i}\right\rbrace_{i=0}^{l}$ with $\vec{r}_{i_0}=\vec{r}_{i_0+l}$. The removing of this loop is equivalent to considering the path $\vec{r}_1,\vec{r}_2,...,\vec{r}_{i_0},\vec{r}_{i_0+l+1},...,\vec{r}_{N}$. By repetition of this procedure, we obtain a loop-less path, which is a LERW. It is well-known that this model belongs to $c=-2$ conformal field theory~\cite{lawler2011conformal}, and also $\kappa=2$ schramm-Loewner evolution (SLE$_{\kappa=2}$) in the continuum limit~\cite{schramm2011scaling}.\\
An important question may arise concerning the effect of the geometry of the metric space on this superdiffusion process. In the geometries of non-integral dimensionality, the properties of critical phenomena change considerably~\cite{dhar1977lattices,dhar1978d}, and its properties depend on the fractal dimension, the order of ramification and the connectivity of the host system~\cite{gefen1980critical}. The LERW in the uncorrelated percolation lattice has been studied in~\cite{daryaei2014loop}, in which it was shown that for the critical host cluster $p=p_c$, the fractal dimension is equal to $D_f\approx 1.22$. To bring the correlations in the diluteness pattern of the host media, one may use a binary model with tunable correlations. The Ising model is a good candidate for this purpose, since it deals with artificial spins ($\sigma$), and the correlations are tuned by an artificial temperature ($T$). The spins play the role of the field of activity-inactivity (diluteness pattern) of the media in the present paper. We consider the majority spin sites, as the active sites through which the random walkers can pass. To this end, we identify the spanning cluster (the set of connected active sites which connects two apposite boundaries), and let the random walkers to have dynamics only over the active area, i.e. the walks is only possible on the active ($s=+1$) sites in the spanning cluster. The sites of the cluster of active area have coordination numbers that is $z_j\equiv \sum_{i\in \delta_i}\delta_{s_i,1}$ in which $\delta_i$ is the set of neighbors of $i$th site and $\delta$ is the Kronecker delta function. This has schematically shown in Fig~\ref{fig:Movement}, for which the coordination number of the central site is three. It is notable that the activity configuration of the media is quenched, i.e. when an Ising configuration is obtained, SAW samples are generated in the resulting dilute lattice. \\
If we show the Ising spins by $s$, then $s=+1$ ($s=-1$) are attributed to the active (inactive) sites, and the temperature is the control parameter which tunes the correlations of the host system. The Ising Hamiltonian is
\begin{equation}
H=-J\sum_{\left\langle i,j\right\rangle}s_is_j-h\sum_{i}\sigma_i, \ \ \ \ \ s_i=\pm 1
\label{Eq:Ising}
\end{equation}
in which $J$ is the coupling constant, $h$ is the magnetic field, $s_i$ and $s_j$ are the spins at the sites $i$ and $j$ respectively, and $\left\langle i,j \right\rangle$ shows that the sites $i$ and $j$ are nearest neighbors. $J>0$ corresponds to positively correlated host system, whereas $J<0$ is for negatively correlated one. The artificial temperature $T$, in addition to being the control parameter of the correlations, controls also the population of the active sites to the total number of sites. This population can also be directly controlled by $h$ which determines the preferred direction of the spins in the Ising model. For $h=0$ (which is the case for the present paper) the model is well-known to exhibit a non-zero magnetization per site $M=\left\langle s_i\right\rangle $ at temperatures below the critical temperature $T_c$. For the 2D regular Ising model at $h=0$ these two transitions occur simultaneously~\cite{delfino2009field}, although it is not the case for all versions of the Ising model, e.g. for the site-diluted Ising model~\cite{najafi2016monte}. \\
We define the Ising model on the $L\times L$ square lattice. Then by solving the Eq.~\ref{Eq:Ising} for $h=0$ an Ising sample at a temperature $T\leq T_c$ is made, and some random walkers start the motion from the boundaries on the largest spanning cluster of the sample as the host media. Some samples of this process have been shown in Fig.~\ref{IsingSamples}. The emphasis of two first figures is on the Ising-correlated site-diluted lattice at two artificial temperatures $T=1.9$ and $T=2.269$. LERWs are evident in these figures. In the two last figures we show the diffusion of particles for two other Ising samples with the mentioned temperatures. The forbidden islands are evident in these figures. The process of random walks is terminated, when they reach one of the boundaries.

\subsection{Numerical details}
We have used the Wolff Monte Carlo method to simulate the system in the vicinity of the critical point to avoid the problem of critical slowing down. Our ensemble averaging contains both random walks averaging as well as Ising-configuration averaging. For the latter case we have generated $2\times 10^3$ Ising uncorrelated samples for each temperature on the lattice size $L=2048$. To make the Ising samples uncorrelated, between each successive sampling, we have implied $L^2/3$ random spin flips and let the sample to equilibrate by $500L^2$ Monte Carlo steps. The main lattice has been chosen to be square, for which the Ising critical temperature is $T_c\approx 2.269$. Only the samples with temperatures $T\leq T_c$ have been generated, since the spanning clusters (active space) are present only for this case. As stated in the previous section the random walkers move only on the active space. The temperatures considered in this paper are $T=T_c-\delta t_1\times i$ ($i=1,2,...,5$ and $\delta t_1=0.01$) to obtain the statistics in the close vicinity of the critical temperature $T_c\simeq 2.269$ and $T=T_c-\delta t_2\times i$ ($i=1,2,...,10$ and $\delta t_2=0.05$) for the more distant temperatures. To equilibrate the Ising sample and obtain the desired samples we have started from the high temperatures ($T>T_c$). For each temperature $2\times 10^6$ LERWs were generated for $2\times 10^3$ Ising samples. We have used the Hoshen-Kopelman~\cite{hoshen1976percolation} algorithm for identifying the clusters in the lattice.

\section{measures and results}\label{NUMDet}

The winding-angle statistics is a very powerful method to extract the key parameters of a two-dimensional critical system, i.e. the diffusivity parameter of the Schramm-Loewner evolution (SLE). The SLE theory aims to describe the interfaces of two-dimensional (2D) critical models via growth processes, and analyzes classifies them to the one-parameter classes represented by $\kappa$. Thanks to this theory, a deep connection between the local properties and the global (geometrical) features of the 2D critical models has been discovered. These non-intersecting interfaces are assumed to have two essential properties, conformal invariance and the domain Markov property~\cite{cardy2005sle}. For more information on the SLE theory see \cite{cardy2005sle,lowner1923untersuchungen}. The relation between the fractal dimension of the curves $D_f\equiv \frac{1}{\nu}$ and the diffusivity parameter ($\kappa$) is $D_f=1+\frac{\kappa}{8}$. Also the CFT/SLE correspondence is satisfied by means of a relation between the central charge of charge of CFT, and the diffusivity parameter of SLE, namely $c=\frac{(6-\kappa)(3\kappa-8)}{2\kappa}$. \\
There are many tests for SLE, most importantly the left passage probability~\cite{najafi2013left,Najafi2015Fokker}, the direct SLE mapping~\cite{Najafi2012Observation,najafi2015observation} and the winding angle statistics~\cite{duplantier1988winding}. The winding angle $\theta$ is the total winding angle of the movement at the end point and a global direction, and is calculated by
\begin{equation}
\text{Var}\left[ \theta\right] = \kappa\log R
\end{equation}
in which $\text{Var}\left[ \theta\right]\equiv \left\langle \theta^2\right\rangle-\left\langle \theta\right\rangle^2$, $\left\langle \right\rangle$ means ensemble averaging, and $R$ is end-to-end distance. Noting that $R$ scales with the trace length (or equivalently the time) $l$ in the form $R\sim l^{\nu}$ (which defines the $\nu$ exponent), one finds that $\text{Var}\left[ \theta\right] = \kappa\nu\log l$. For LERW in $T=0$ (the regular lattice), this slope is $1.6$. Also the fractal dimension of a curve (in the box-counting scheme) is defined by the relation $l(L)\sim L^{D_f}$, in which $L$ is the linear size of a box which contains a curve of length $l$.\\

\begin{figure*}
	\centering
	\begin{subfigure}{0.49\textwidth}\includegraphics[width=\textwidth]{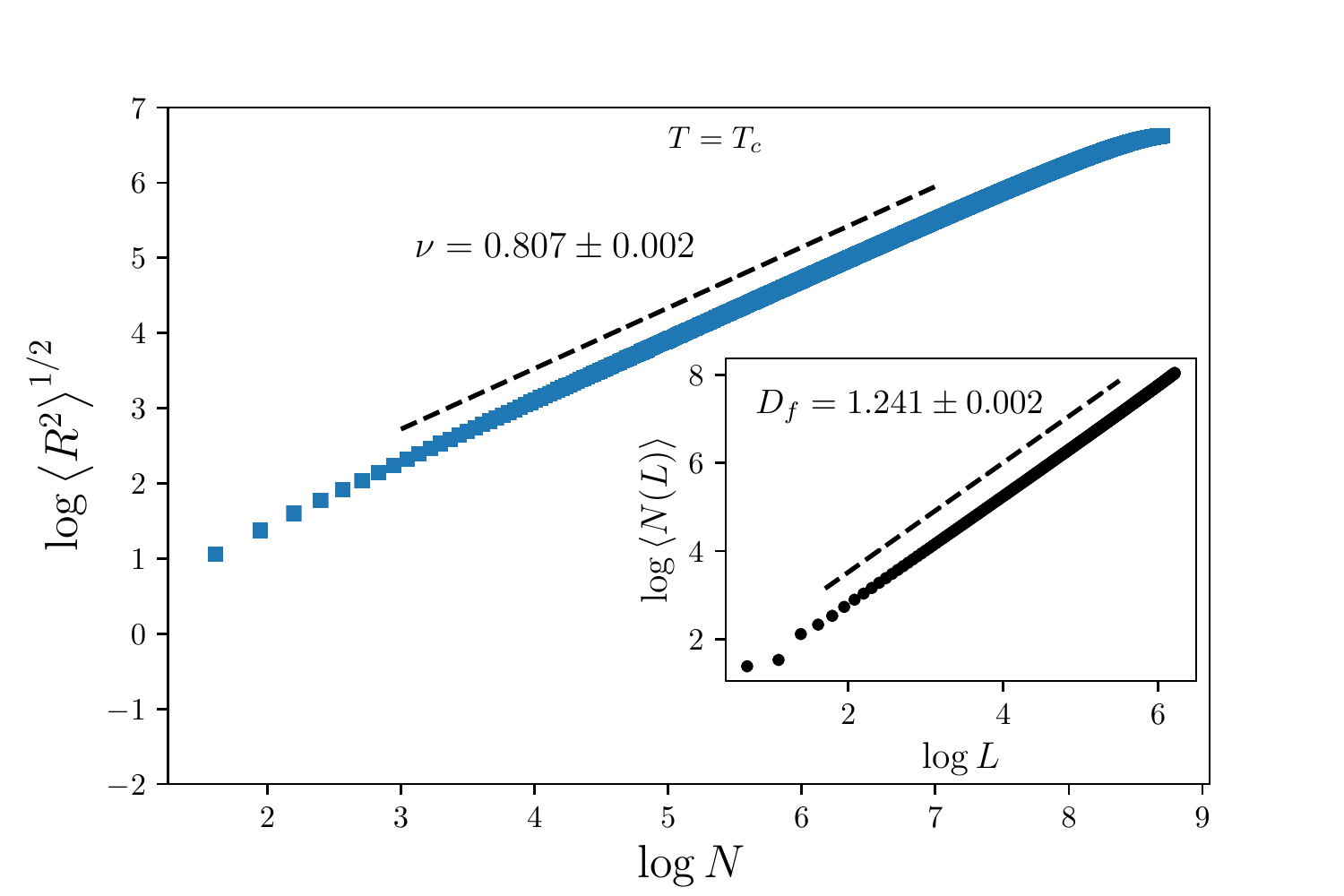}
		\caption{}
		\label{fig:r_Tc}
	\end{subfigure}
	\begin{subfigure}{0.49\textwidth}\includegraphics[width=\textwidth]{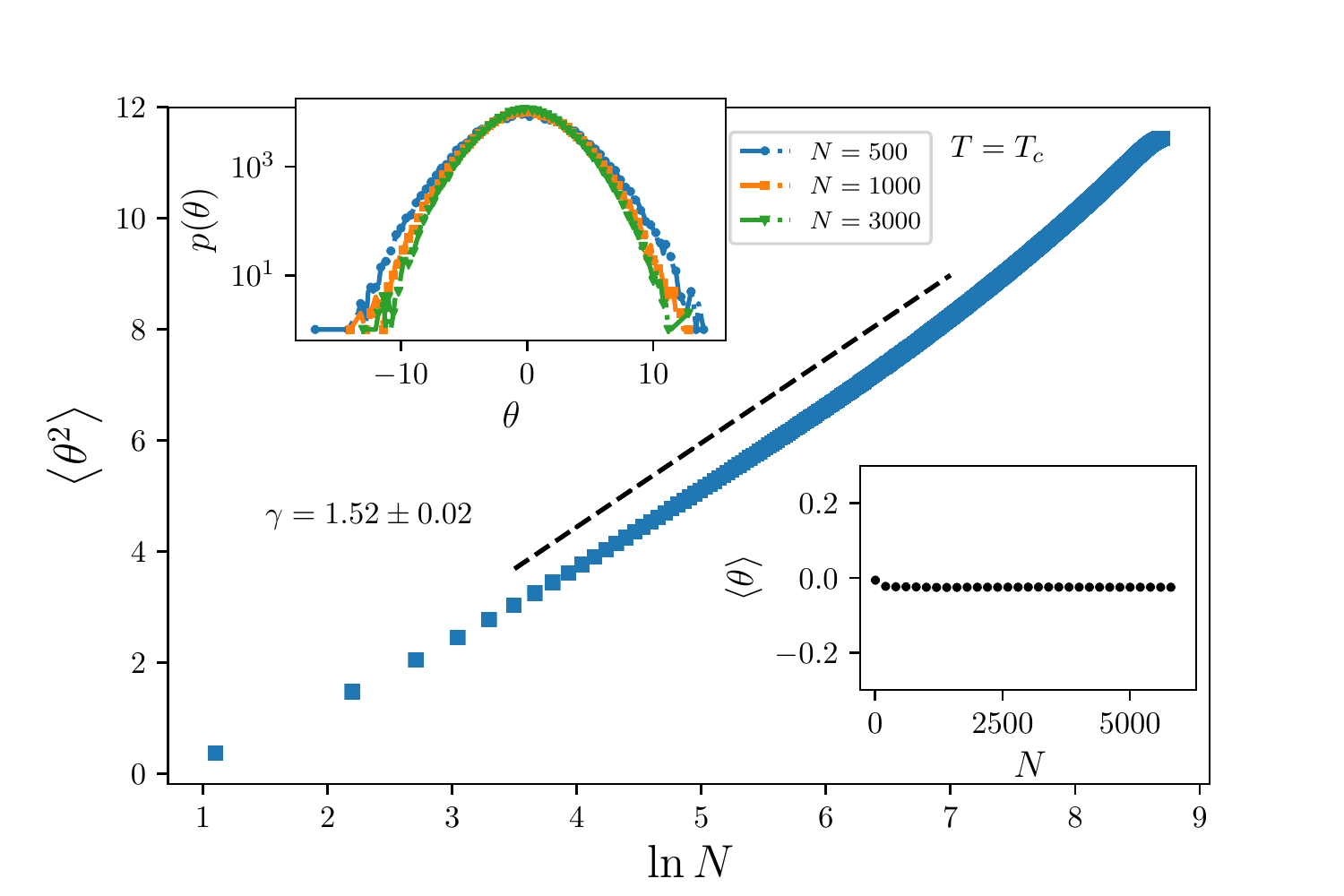}
		\caption{}
		\label{fig:WA-Tc}
	\end{subfigure}
	\caption{(Color online): (a)  $\log_{10}\left\langle R^2\right\rangle^{\frac{1}{2}}$ in terms of $\log_{10}N$ for $T=T_c$. Inset: $\log_{10}\left\langle N(L)\right\rangle$ in terms of $\log_{10}L$. (b) $\left\langle \theta^2\right\rangle $ in terms of $\ln N$ with the slope $\gamma=1.52\pm 0.02$ for $T=T_c$. Upper inset: the distribution function of $\theta$ for various rates of $N$. Lower inset: $\left\langle \theta\right\rangle$ in terms of $N$ which is identical to zero.}
	\label{fig:Tc}
\end{figure*}
Now let us consider the critical lattice, i.e. $T=T_c$. In this case the exponents exhibit deviation with respect to the regular lattice. In the Fig.~\ref{fig:Tc} the exponents $\nu$, $D_f$, and $\kappa$ have been calculated. Figure \ref{fig:r_Tc} reveals that $\nu=\frac{1}{D_f}$ is valid at the critical lattice. We note that the $\nu$ exponent $0.9\%$ changes from its value for regular lattice, i.e. $\nu_{\text{LERW}}=0.8$.\\
The statistics of $\theta$ has been presented in Fig.~\ref{fig:WA-Tc}. We see that the distribution of $\theta$ ($p(\theta)$) remains Gaussian at $T=T_c$, and its variance grows with $N$ (the number of walkers steps). The variance of $\theta$ grows linearly with $\ln N$ and $\left\langle \theta\right\rangle\approx 0$. In this figure $\gamma=\kappa\nu$ is the slope of the linear graph. This guarantees that the trace of walkers is conformal invariant, i.e. they are SLE traces. The results have been gathered in TABLE~\ref{tab:nu}.
\begin{table}
	\begin{tabular}{c | c c c c}
		\hline & $\kappa_{T_c}$ & $\kappa_{T_c}(D_f)$ & $\nu_{T_c}$ & $D_F^{T_c}$ \\
		\hline direct & $1.89\pm 0.05$ & $1.93\pm 0.02$ & $0.807\pm 0.002$ & $1.241\pm 0.002$ \\
		\hline
	\end{tabular}
	\caption{The exponents of LERW at $T=T_c$. $\kappa_{T_c}$ has been obtained by winding angle analysis, and $\kappa_{T_c}(D_f)$ has been obtained by the relation $\kappa=8(D_f-1)$. The $\nu_{T_c}$ and $D_F^{T_c}$ have been obtained using the relation $R\sim l^{\nu}$ and the box-counting method.}
	\label{tab:nu}
\end{table}

\subsection{The general behavior in terms of $T$}\label{offcritical}

\begin{figure*}
	\centering
	\begin{subfigure}{0.49\textwidth}\includegraphics[width=\textwidth]{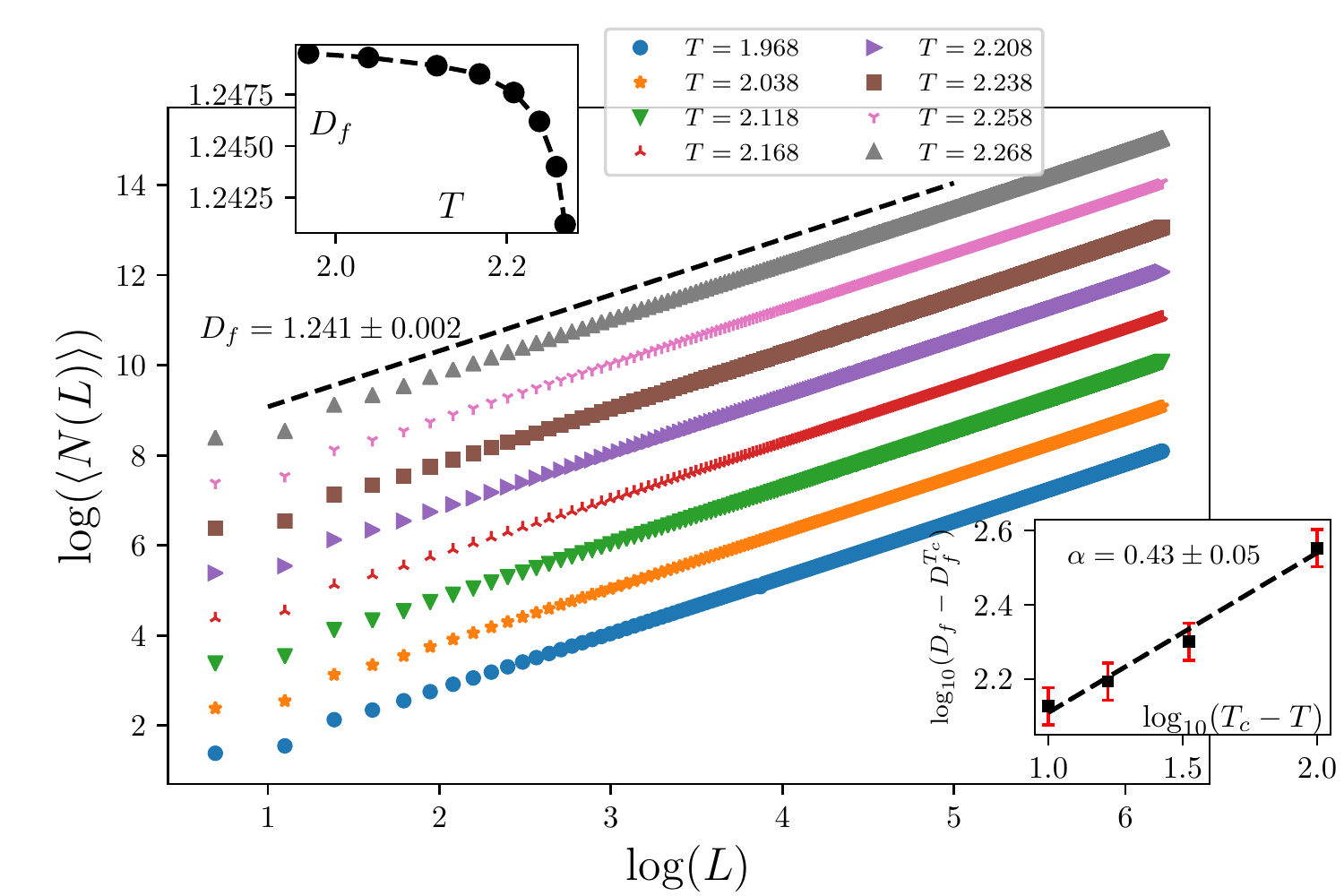}
		\caption{}
		\label{fig:BC}
	\end{subfigure}
	\begin{subfigure}{0.49\textwidth}\includegraphics[width=\textwidth]{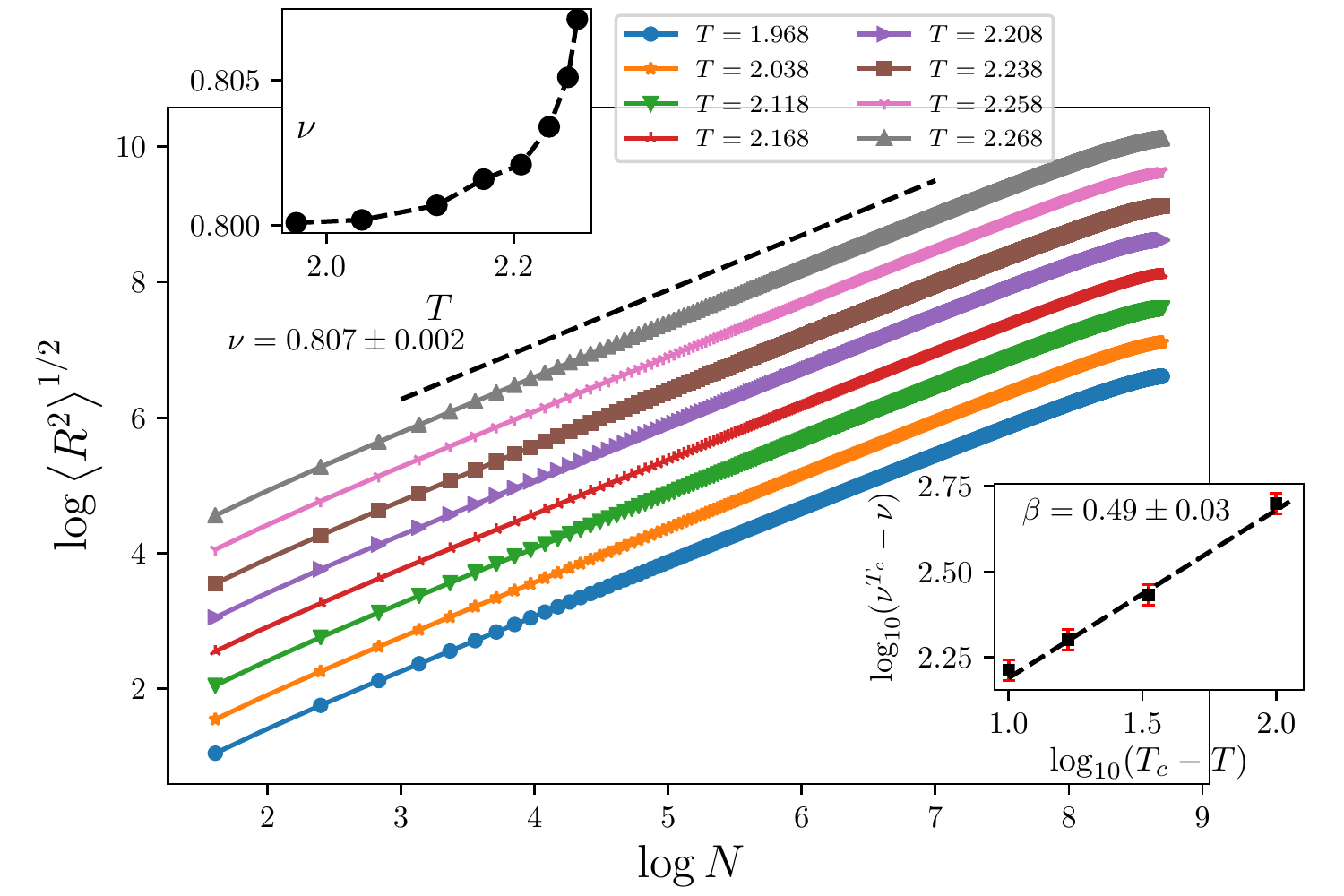}
		\caption{}
		\label{fig:rms}
	\end{subfigure}
	\begin{subfigure}{0.49\textwidth}\includegraphics[width=\textwidth]{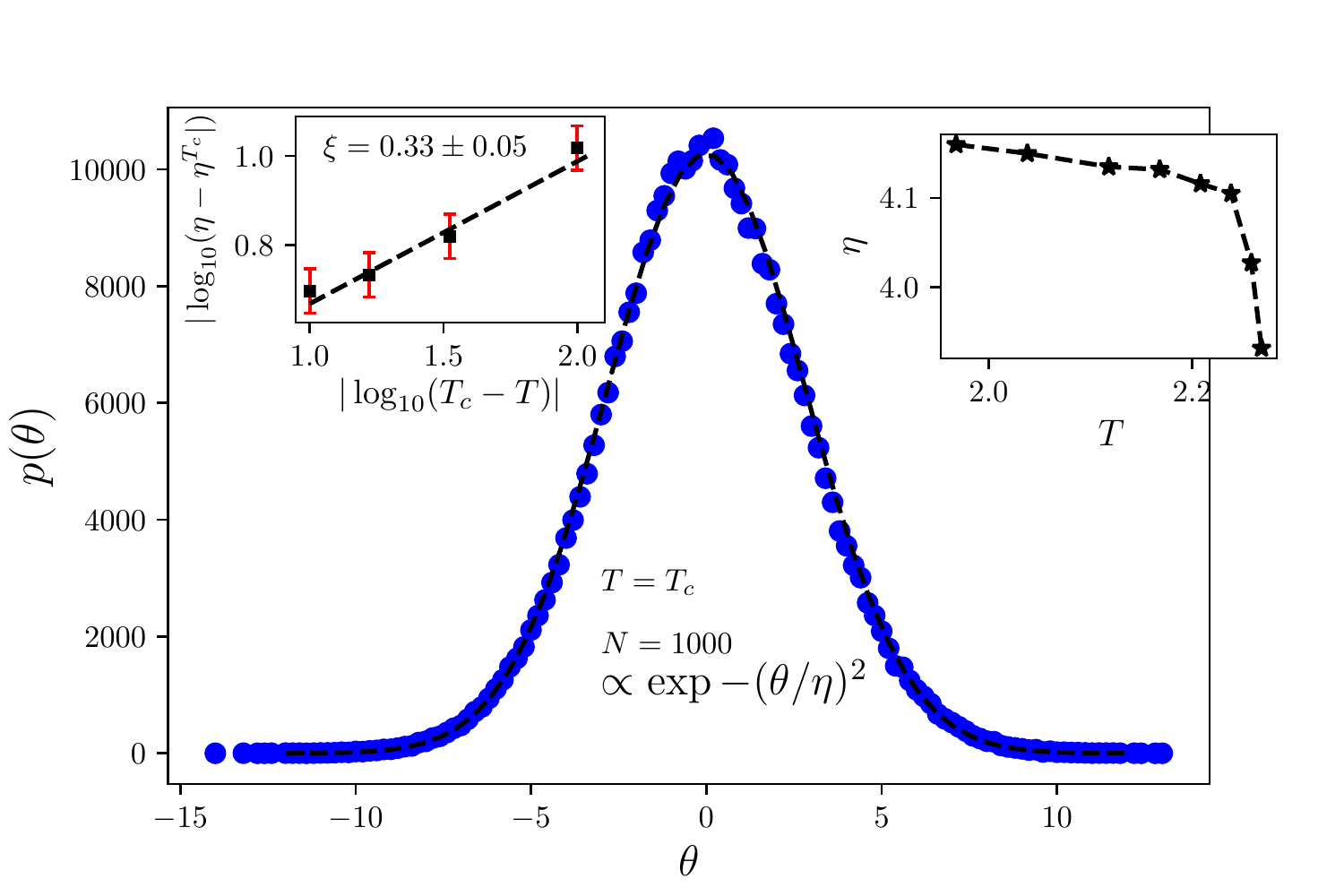}
		\caption{}
		\label{fig:theta}
	\end{subfigure}
	\begin{subfigure}{0.49\textwidth}\includegraphics[width=\textwidth]{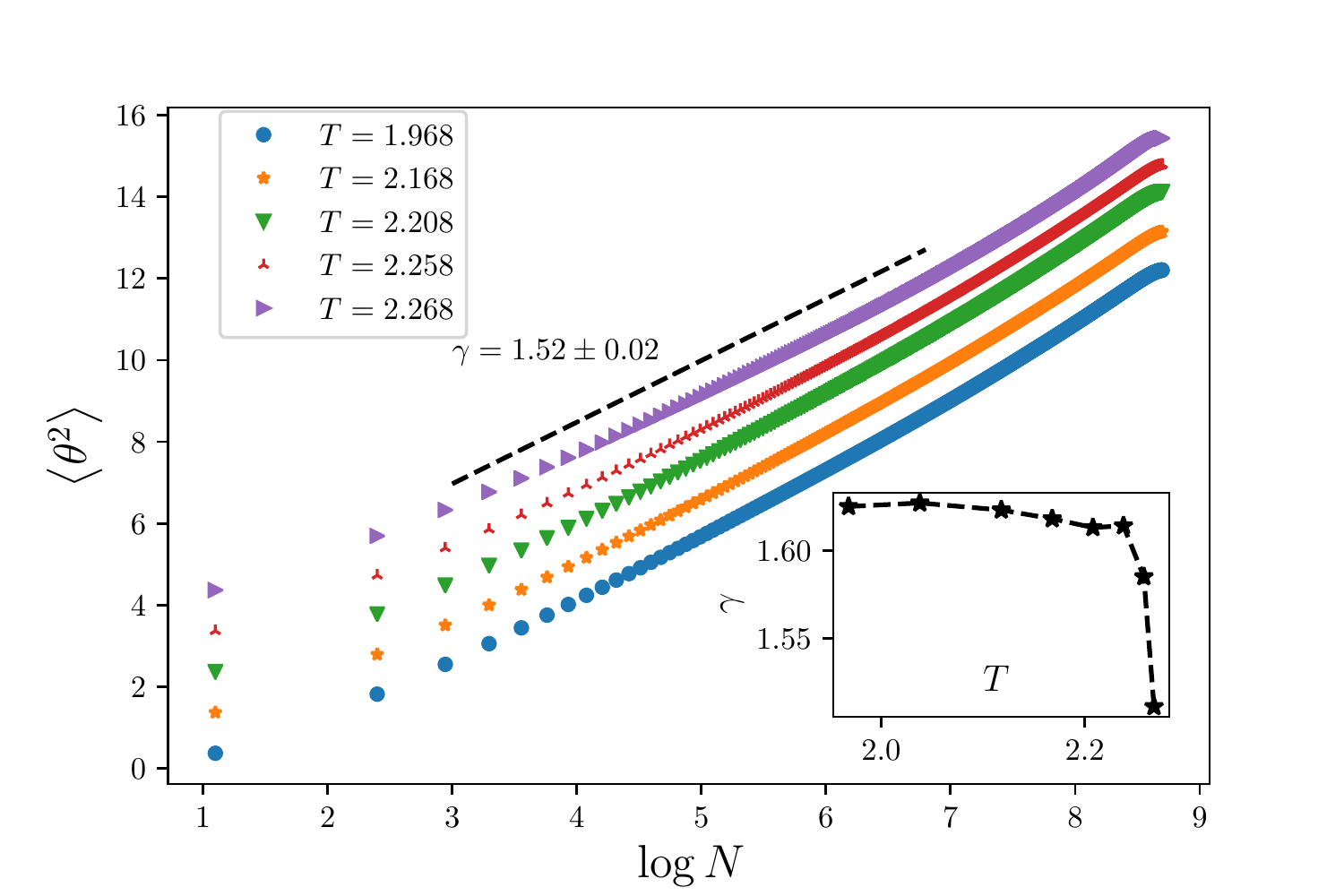}
		\caption{}
		\label{fig:WA}
	\end{subfigure}
	\caption{(Color online) (a) $\log_{10}\left\langle N(L)\right\rangle$ in terms of $\log_{10}L$ for various rates of temperature. Upper inset: $D_F$ in terms of $T$, Lower inset: power-law behavior of the fractal dimension. (b) $\log_{10}\left\langle R^2\right\rangle^{\frac{1}{2}}$ in terms of $\log_{10}N$ for various rates of temperature. Upper inset: the $\nu$ exponent in terms of $T$, Lower inset: power law behavior of $\nu$. (c) The distribution of the winding angle $\theta$ for $N=1000$ and $T=T_c$, which is fitted to $\exp-\left(\theta/\eta\right)^2$. Insets: the power-law and ordinary behaviors of the $\eta$ in terms of $T$. (d) $\left\langle \theta^2\right\rangle $ in terms of $\ln N$ with the slope $\gamma$ which is $T$-dependent. Inset: $\gamma$ in terms of $T$.}
	\label{fig:Off-Tc}
\end{figure*}
We have seen that for the critical lattice ($T=T_c$), the exponents are displaced with respect to the regular lattice. In this case the effective dimension of the system is $\bar{d}=\frac{187}{96}\simeq 1.948$~\cite{duplantier1989exact}. An equally worthy problem is the diffusion process in the super-critical systems, i.e. $T<T_c$ in which the $\bar{d}(T_c)<\bar{d}(T)<2$ (the effective dimension of the host media) varies. In this case, the trace of particles in the diffusion process should be denser (due to more available active regions) resulting to larger fractal dimensions and equivalently lower $\nu$s. This variation is not surely predetermined. Importantly, one may expect some power-law behaviors, due to the criticality of the system in this regime. These power-law behaviors in the geometrical observables arises from the scaling laws of the correlation length $\zeta(T)$ of the Ising model as the host of the system. This characteristic length tends to infinity at $T=T_c$ in the thermodynamic limit. \\
In Fig.~\ref{fig:Off-Tc} we have shown the results for this case. The most important results are sketched in Figs.~\ref{fig:BC} and~\ref{fig:rms}. It is realized from these figures that $D_f(T)-D_f(T_c)\sim t^{\alpha}$ and $D_f(T)-D_f(T_c)\sim t^{\beta}$, in which $t\equiv\frac{T_c-T}{T_c}$, and $\alpha=0.43\pm 0.05$, and $\beta=0.49\pm 0.03$. We note that the correlation length of the Ising model scales as $\zeta\sim t^{-1}$. Approximating $\alpha$ and $\beta$ by their the closest fractional value ($\frac{1}{2}$), one finds that $D_f(T)-D_f(T_c)\sim \frac{1}{\sqrt{\zeta(T)}}$. This result has been found in many statistical models on the Ising-correlated lattices, like Gaussian free field~\cite{cheraghalizadeh2018gaussian}, self-avoiding walk~\cite{cheraghalizadeh2018self}, and sandpile models~\cite{cheraghalizadeh2017mapping}.\\
The winding angle statistics also yield some information concerning the diffusivity parameter $\kappa$. In Figs.~\ref{fig:theta} and~\ref{fig:WA} we have shown the same functions as the Fig.~\ref{fig:WA-Tc}. It is seen that the variance of $\theta$ ($\sim\eta$) for $N=1000$ increases with decreasing $T$. This means that the traces tend to twist more in lower temperatures. This is compatible with the obtained fractal dimensions, i.e. more fractal dimensions are equivalent to denser traces which apparently should twist more than the dilute system. $\gamma$ increases also with decreasing $T$ which has been shown in the inset of Fig.~\ref{fig:WA}. Extending the relation $\gamma=\kappa\nu=8\nu(D_f-1)$ to off-critical temperatures, we expect that this exponent does not show the power-law behaviors.

\section*{Discussion and Conclusion}
\label{sec:conc}
In the present paper we have considered the effects of the correlations in the dilute pattern in the system, on the properties of the superdiffusion process with $\nu=\frac{4}{5}$. Loop-erased random walk has been employed to simulate the particle's motion in this process. We also have modeled the forbidden regions into which the particles are not allowed to enter by the Ising model with an artificial temperature $T$, that tunes the host system correlations. The spins of the Ising model play the role of occupation field in the real system. We have observed that the results change considerably with respect to the un-correlated host system. Importantly at the critical temperature $T=T_c$, the exponent of the end-to-end distance becomes $\nu=0.807\pm 0.002$. This shows that the resulting particle traces become more dilute in this case. The winding angle test also shows that the variance of the angle $\theta$ behaves linearly with $\ln R$, which yields the diffusivity parameter of the SLE theory $\kappa=1.89\pm 0.05$. The other predictions of the conformal invariance are also satisfied, i.e. the fractal dimension has the relation with $\nu$ and $\kappa$ as $D_f=\frac{1}{\nu}=1+\frac{\kappa}{8}$.\\
In the off-critical regime $T<T_c$, in addition to the ordinary power-law behaviors of the statistical observables (which gives rise to some critical exponents), some power-law behaviors are also seen for the exponents in terms of $t\equiv\frac{T-T_c}{T_c}$. Importantly we have observed that $D_f(T)-D_f(T_c)\sim t^{0.43\pm 0.05}$. Using the well-known result of the Ising model for the correlation length $\zeta(T)\sim t^{-1}$ in the thermodynamic limit, and approximating the exponent with $\frac{1}{2}$, we see that $D_f(T)-D_f(T_c)\sim \zeta^{-\frac{1}{2}}$, which has been observed in many other statistical models on the Ising-correlated lattices. The full exponents have been listed in TABLE~\ref{tab:off}.

\begin{table}
	\begin{tabular}{c | c c c}
		\hline exponent & definition & value & closest fractional value \\
		\hline $\alpha$ & $D_f(T)-D_f{T_c}\sim t^{\alpha}$ & $0.43\pm 0.05$ & $\frac{1}{2}$ \\
		\hline $\beta$ & $\nu(T)-\nu{T_c}\sim t^{\beta}$ & $0.49\pm 0.02$ & $\frac{1}{2}$ \\
		\hline $\xi$ & $\eta(T)-\eta(T_c)\sim t^{\xi}$ & $0.33\pm 0.05$ & $\frac{1}{3}$ \\
		\hline
	\end{tabular}
	\caption{The off-critical exponents of the problem, with their definitions, and the suggested fractional value.}
	\label{tab:off}
\end{table}

\bibliography{refs}

\end{document}